\documentclass[aps,pra,twocolumn,superscriptaddress]{revtex4}

\usepackage{dcolumn}
\usepackage[utf8]{inputenc}
\usepackage{lmodern}
\usepackage[T1]{fontenc}
\usepackage[english]{babel}
\usepackage[dvips]{graphicx}
\usepackage{enumerate,amsthm,amsmath,amssymb,color,graphicx,bbm,float}
\usepackage{bm}
\usepackage{mathtools}
\usepackage{natbib}
\usepackage{subfigure}
\usepackage{synttree}
\usepackage{overpic}

\usepackage{epstopdf}

\usepackage{tikz}
\usetikzlibrary{arrows}

\newcommand{\be}{\begin{equation}}
\newcommand{\ee}{\end{equation}}

\def\multiset#1#2{\ensuremath{\left(\kern-.3em\left(\genfrac{}{}{0pt}{}{#1}{#2}\right)\kern-.3em\right)}}

\begin{document}

\preprint{APS/123-QED}

\title{High Bit Rate Continuous-Variable Quantum Key Distribution}

\author{Paul Jouguet}\affiliation{SeQureNet, 23 avenue d'Italie, 75013 Paris, France}
\author{David Elkouss}\affiliation{Departamento de An\'alisis Matem\'atico, Universidad Complutense de Madrid}
\author{S\'ebastien Kunz-Jacques}\affiliation{SeQureNet, 23 avenue d'Italie, 75013 Paris, France}

\date{\today}

\begin{abstract}
%%%% Changed DE
Here, we demonstrate that a practical Continuous Variables Quantum Key Distribution (CVQKD) protocol relying on the Gaussian modulation of coherent states features secret key rates that cannot be achieved with standard qubit Discrete Variables (DV) QKD protocols. Notably, %specifically
we report for the first time a practical postprocessing that allows to extract more than one bit of secret key per channel use. 
%%%% Changed DE
%We study the performance of practical reconciliation schemes for Continuous Variables Quantum Key Distribution (CVQKD). While complex post-processing is required in order to process continuous information, CVQKD allows the participants of the protocol to extract more than one bit of secret key per pulse, which is not possible with standard qubits Discrete Variables (DV) QKD protocols. Here, we study the parameters regime where a practical CVQKD protocol relying on the Gaussian modulation of coherent states and homodyne detection  %\cite{GG:prl02} 
%features secret key rates that cannot be achieved with DVQKD protocols. 
%%%%%%%%%%%%%%%
%%Notably, we report for the first time a practical postprocessing that allows to extract more than one bit of secret key per channel use. 
%%%%%%%%%%%%%%%
%We apply these results to a state of the art CVQKD system whose hardware repetition rate is 1 MHz, which is significantly lower than its DVQKD 1 GHz counterpart, and make projections about future achievable secret key rates.
\end{abstract}

%\pacs{03.65.Ud, 03.67.-a, 03.67.Dd}
\maketitle

\section{Introduction}
QKD \cite{SBC:rmp09} has been the most studied quantum information technology primitive for the past twenty years. In a practical QKD protocol, Alice and Bob can extract an arbitrary amount of secret key using an untrusted physical channel (also called quantum channel), provided a few minimum assumptions such as they have access to a public authenticated channel. Contrary to classical cryptographic primitives whose security can be established only against some restrictive classes of eavesdroppers, QKD keys are secure in the information-theoretic sense even against an eavesdropper with unlimited computational resources or with undisclosed cryptanalytic knowledge.

In DVQKD protocols, the information is encoded on discrete values, such as the phase or the polarization of single photons, and detection is done using single photon detectors. CVQKD protocols employ continuous or discrete modulations \cite{WPG:rmp12} of the quadratures of the electromagnetic field. CVQKD setups rely on a coherent detection (homodyne or heterodyne) between the quantum signal and a classical reference signal called the local oscillator, and their implementation requires only standard telecom components. They are compatible with Wavelength Division Multiplexing \cite{QZQL10} which greatly eases their deployment into telecommunication networks. In the early history of CVQKD, this technology was expected to achieve higher secret key rates than DVQKD protocols thanks to the possibility of encoding more than one bit per pulse. However, the secure distance of the most common CVQKD protocol \cite{GG:prl02}, which consists in a Gaussian modulation of coherent states in the phase space and a homodyne detection of any of two orthogonal quadratures of the field at random, was limited to 25 km \cite{LBG:pra07} for a long time because of the lack of efficient error-correction procedures at low signal-to-noise ratios. This problem was solved thanks to the multidimensional reconciliation technique proposed in \cite{LAB:pra08} together with the design of high efficiency error correcting codes in \cite{JKL:pra11} and significantly extended the secure distance of CVQKD to about 80 km \cite{JKL+:natphoton13}. However, multidimensional protocols are limited to one bit per pulse.

%%%% Changed DE
In this paper, we exhibit high-efficiency error correcting codes for the Additive White Gaussian Noise Channel (AWGNC). In the high signal-to-noise ratios (SNRs) regime, it allows us to go beyond previous achievable secret key rates \cite{JKL+:natphoton13} with CVQKD systems and extract more than one bit of secret key per channel use; a rate impossible to attain, even in principle, with qubit DVQKD systems.% \cite{PDL+:apl2014}. 
%%%% Changed DE
%In this paper, we exhibit high-efficiency error correcting codes for the Additive White Gaussian Noise Channel (AWGNC). In the high signal-to-noise ratios (SNRs) regime, it allows us to go beyond previous achievable secret key rates \cite{JKL+:natphoton13} with CVQKD systems and extract more than one bit of secret key per channel use which is not possible with state-of-the-art DVQKD systems \cite{PDL+:apl2014}. 

In Section \ref{sec:error_correction_cv}, we explain the links between the secret key rate and error correction in CVQKD and review previous work on error correction for both DVQKD and CVQKD. In Section \ref{sec:slice_reconciliation} we detail the principle of Slice Reconciliation, which is a technique that can be used to reconcile non-binary elements, and study its practical performance in the specific case of the distribution of Gaussian elements. Finally, we show in Section \ref{sec:application} the consequences of these developments on the performance of the Gaussian protocol over short distances with a state of the art CVQKD system and make projections about future achievable secret key rates.

\section{Error Correction with Continuous Variables}
\label{sec:error_correction_cv}

\subsection{Secret Key Rate and Error Correction}

In any QKD protocol (either DV or CV), after some quantum states are exchanged on a quantum channel, an error correction mechanism is used to make Alice and Bob share some common data. There  are two usual cases: either Bob corrects its errors with respect to Alice in the \emph{direct reconciliation} scenario; or Alice corrects its errors with respect to Bob in the \emph{reverse reconciliation} scenario. In these two cases, the party performing error correction does so using additional data revealed by the other party through a noiseless, classical channel.

The final secret key size generated by a QKD experiment therefore depends on three quantities: the raw common data after error correction, the amount of information that was revealed during the error correction phase, and an upper estimate of the amount of information gained by the attacker through its interaction with the quantum channel. The latter quantity is a result of the security proof of the considered protocol, and is the information that the attacker Eve has in common with Alice in the direct reconciliation case and with Bob in the reverse reconciliation case. In the case of CVQKD, the measurement of information used is the Holevo information and the direct (resp. reverse) quantities are denoted by $\chi_{AE}$ (resp. $\chi_{BE}$). The relevant quantity to take into account for the amount of information revealed because of error correction is the mutual (Shannon) information between Alice and Bob, $I_{AB}$. A perfect error-correction scheme is able to retrieve all of $I_{AB}$, that is, the amount of common 
information after error correction substracted of the amount of auxiliary data revealed to perform the error correction is equal to $I_{AB}$; a practical scheme will extract only an amount of information $\beta I_{AB}$ with $\beta < 1$. Overall, the final amount of secret key produced by a QKD protocol is $\beta I_{AB} - \chi_{AE}$ with direct reconciliation and $\beta I_{AB} - \chi_{BE}$ with reverse reconciliation.

In the case of Gaussian modulated coherent-state CVQKD  \cite{GG:prl02}, the channel parameters enabling to bound the information obtained by Eve are the line transmission $T$ and the noise added by Eve on the quantum channel or \emph{excess noise} $\xi$. When there is no excess noise, one has for any line transmission $T$, $\chi_{BE} < I_{AB}$: some secret key can be produced at any distance using reverse reconciliation with a perfect error correction scheme, or a ``sufficiently good'' scheme such that $\beta I_{AB} - \chi_{BE} > 0$. Using direct reconciliation however, $\chi_{AE} < I_{AB}$ only when losses are lower than 3 dB ($T>0.5$), therefore a direct reconciliation scheme can be used only for short distances.

In coherent-state Gaussian CVQKD \cite{GG:prl02}, the error correction schemes used also depend on the SNR of the data to correct. For this protocol, the error-correction scenario is a bit unusual since both the signal and the noise are Gaussian, which is not a well studied scenario outside the field of QKD. The error correction must also be paired with an algorithm to extract identical bits out of highly correlated Gaussian values. The efficiency factor $\beta$ is typically highly sensitive to the SNR of the system; historical CVQKD systems used reasonably high SNRs because of this. The coherent-state Gaussian protocol is the CVQKD protocol whose security has been studied the most because it features higher secret key rates than protocols that employ discrete modulations and can be implemented with standard components contrarily to squeezed-state protocols. However, in contrast to DVQKD, specific error correction techniques need to be designed to deal with non-binary key elements. Furthermore, the error correction schemes used also depend on the SNR of the data to correct.

\subsection{Previous work}
The first reconciliation protocols were ad-hoc constructions targeting DVQKD.  Among these early proposals, Cascade \cite{BS:eurocrypt93} stands out as a very simple protocol with reasonably high efficiency. Its principal defect is that it is extremely interactive. However, a recent implementation of Cascade \cite{PT13} shows that, provided that a dedicated classical communications line is available, a high throughput is achievable. 

In contrast to these protocols, most recent work in DVQKD has focused in applying capacity-approaching one-way error correcting codes for reconciliation. For instance, large length ($10^6$) low-density parity-check (LDPC) codes can be used to approach the theoretical limits \cite{ELAB:isit09}. These results only hold for large lengths, recently explicit fundamental one-way limits have been stressed in \cite{TMPE:isit14} as a function of the length and the target frame error rate (FER). However, a combination of error correcting codes with a few rounds of interactivity allows to bypass these limitations while maintaining a high throughput \cite{MEM:sr13}.

As regards CVQKD, specific error correction techniques need to be designed to deal with non-binary key elements. In \cite{VAN04}, Slice Error Correction (SEC) was proposed to extract mutual information out of any correlated variables, either discrete or continuous. SEC uses interactive error correcting codes whose efficiency is suboptimal as pointed in \cite{BLO05}. MultiLevel Coding / MultiStage Decoding (MLC / MSD) are standard coded modulation techniques that were applied to CVQKD reconciliation in \cite{BLO05} and \cite{LBG:pra07}. They feature higher efficiency than SEC for SNR between 1 and 15 but their efficiency drops quickly for SNR below 3. In CVQKD, achieving long distances requires to maintain a high reconciliation efficiency for low SNRs. This is why the secure distance was limited to 25 km \cite{LBG:pra07} until the multidimensional reconciliation scheme was proposed in \cite{LAB:pra08}. This scheme encodes the information in binary variables which allows us to deal with a Binary Input (BI) AWGNC instead of the usual AWGNC. Since low-rate high-efficiency multi-edge LDPC codes can be designed for this channel \cite{JKL:pra11}, the achievable secure distance for CVQKD with a Gaussian modulation can be considerably extended. In \cite{LAB:pra08}, high efficiency with a SNR of 0.5 allowed to extend the secure distance to about 50 km while LDPC codes specifically designed for SNRs as low as 0.03 \cite{JKL:pra11} were used to demonstrate the exchange of secure keys at 80 km \cite{JKL+:natphoton13}. Finally, since achieving high efficiencies requires intensive iterative decoding for LDPC codes, the use of Graphic Processing Units (GPUs) \cite{LBG:pra07, JK:QIC13} for LDPC decoding or the use of polar codes \cite{JK:QIC13} which feature a high speed decoder on Central Processing Units (CPUs) have been investigated.

\section{Slice Reconciliation}
\label{sec:slice_reconciliation}
\subsection{Principle}
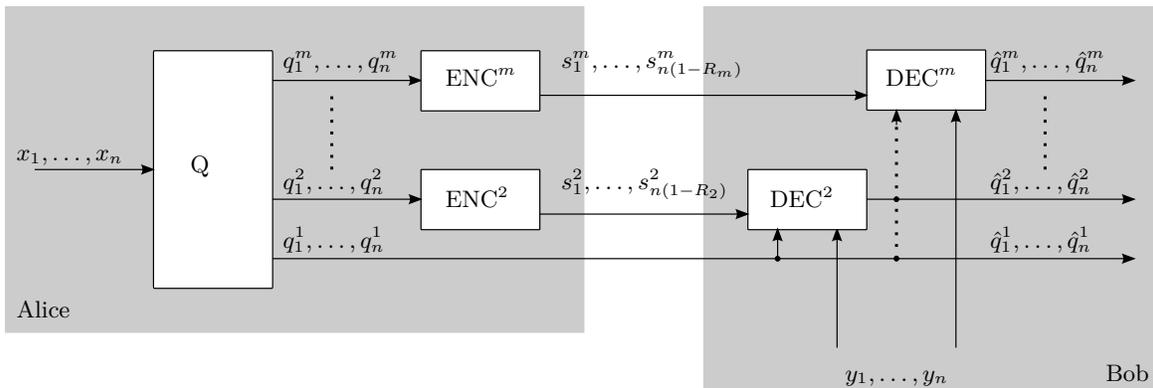
\begin{figure*}[!t]
\centering
\begin{overpic}[scale=1.4,unit=1mm]
{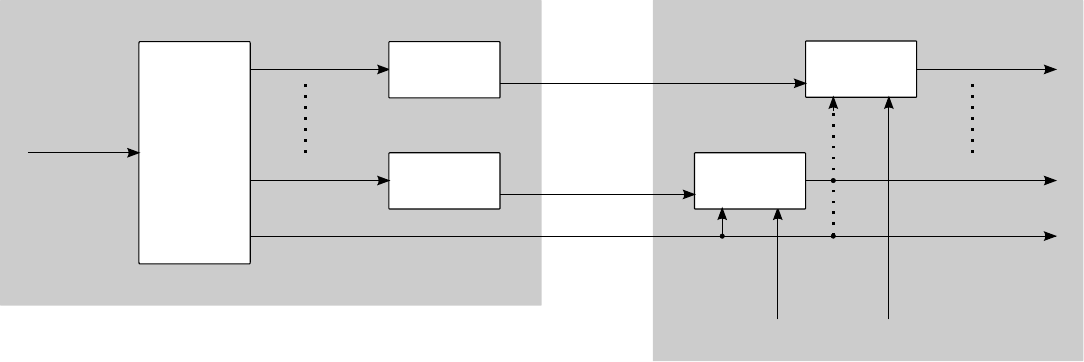}
\put(1,20){$x_1,\hdots,x_n$}
\put(24,12.5){$q_1^1,\hdots,q_n^1$}
\put(24,17.5){$q_1^2,\hdots,q_n^2$}
\put(24,28){$q_1^m,\hdots,q_n^m$}
\put(48,17.5){$s_1^2,\hdots,s^2_{n(1-R_2)}$}
\put(48,28){$s_1^m,\hdots,s^m_{n(1-R_m)}$}
\put(85,12.5){$\hat q_1^1,\hdots,\hat q_n^1$}
\put(85,17.5){$\hat q_1^2,\hdots,\hat q_n^2$}
\put(85,28){$\hat q_1^m,\hdots,\hat q_n^m$}
\put(16,19){Q}
\put(72.5,1){$y_1,\hdots,y_n$}
\put(38,16){$\textrm{ENC}^2$}
\put(38,26.5){$\textrm{ENC}^m$}
\put(66,16){$\textrm{DEC}^2$}
\put(76,26.5){$\textrm{DEC}^m$}
\put(95,1){Bob}
\put(1,6.5){Alice}
\end{overpic}
\caption{Schematic representation of the slice protocol for direct reconciliation. First the input source is quantized into an $m$-bit source. Then each of the $m$ sources is encoded and sent to Bob. In the figure the first slice is transmitted unencoded. The decoder takes as side information its own source and with the $m$ encoded sources produces an estimate of the quantized source.}
\label{fig:protocol}
\end{figure*}

Slice reconciliation was introduced in \cite{VAN04} as a reconciliation scheme for non-binary sources using binary error correcting codes. It works in two steps (see Fig. \ref{fig:protocol} for a schematic description of the protocol). The first step consists in choosing a set of $m$ slice functions $S_1,..S_m:\mathbb R\rightarrow \{0,1\}$ that take the source to binary values. Together the $m$ functions can be regarded as a quantizing function $Q:\mathbb R\rightarrow \{0,1\}^m$ that transforms the continuous Gaussian source into an $m$ bit source. However by the data processing inequality $I(Q(X);Y)\leq I(X;Y)$ (or equivalently $I(X;Q(Y))\leq I(X;Y)$ for RR). That is, there is an inherent inefficiency associated with the discretization of the source. For any fixed number of bits $m$ we can optimize the secret key rate by finding the function that maximizes $I(Q(X);Y)$ ($I(X;Q(Y))$ for RR). This problem of designing a discretization function that maximizes a mutual information criterion was described in \cite{CV:itw03}.

\begin{figure}[h!]
\centering
\begin{overpic}[scale=1.8,unit=1mm]{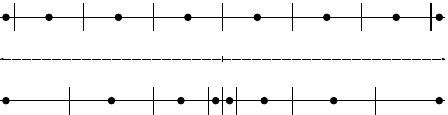}
\put(-6,14){\small{$-\infty$}}
\put(50,14){\small{$0$}}
\put(100,14){\small{$\infty$}}
\end{overpic}
\caption{Two examples of quantizers dividing the real line in $2^3$ intervals. The figure on the top shows a quantizer with constant step, the figure on the bottom shows a quantizer with optimized interval length.}
\label{fig:realline}
\end{figure}

\begin{figure}
\centering
 \includegraphics[width=81mm]{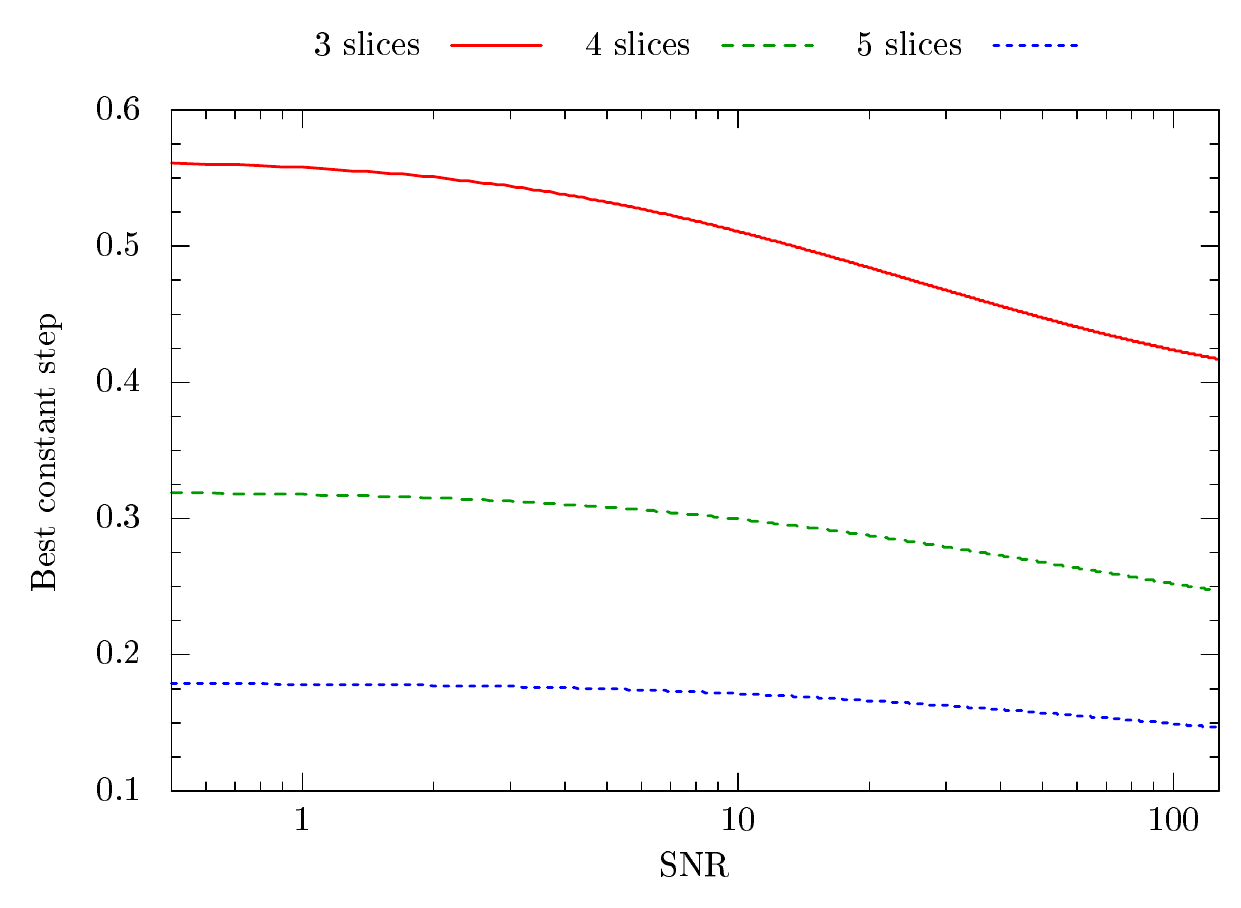}
  \caption{(Color online) Evolution of the value of the constant step giving the best quantization efficiency with respect to the SNR. Solid red line corresponds to 3 slices (8 intervals), dashed green line corresponds to 4 slices (16 intervals), dotted blue line corresponds to 5 slices (32 intervals).}
   \label{figure:bestConstantStep}
\end{figure}

We consider here two different slice constructions (see Fig. \ref{fig:realline}). Both of them divide the real line into $2^m$ disjoint intervals and take the Gaussian source to the ($m$ bit) index of the interval. In the first slice construction the intervals are defined by $2^m-1$ equally spaced points. Finding the function that maximizes the mutual information reduces to optimizing over a single degree of freedom. We report in Fig. \ref{figure:bestConstantStep} the evolution of the value of the constant step giving the best quantization efficiency with respect to the SNR for $m=3$ to $m=5$. The second construction chooses freely the $2^m$ intervals. In this case, finding the optimal function is an optimization problem with $2^m-1$ degrees of freedom. We can see in Fig. \ref{figure:quantizationEfficiency} that this more complex construction does not improve much the quantization efficiency obtained with the first construction. This is why we used the first construction in practice to obtain the reconciliation efficiencies reported in Table \ref{tab:efficiencies}.

The second step deals with sending an encoding of $Q(X)$ to Bob (resp. $Q(Y)$ to Alice in RR) such that he can infer $Q(X)$ (resp. $Q(Y)$ in RR) with high probability. This is a problem that can be readily tackled with coding techniques. In particular, slice reconciliation uses a multilevel coding scheme \cite{BLO05}. Each of the $m$ slices is encoded independently as the syndrome of an error correcting code with rate $R_i$ ($1\leq i\leq m$). If the information rates are chosen appropriately the decoder can recover each of the slices using its own source as side information. The rate of the encoding is upper bounded by the capacity of the associated channel. However, this bound can only be reached in the limit of asymptotically large codes, in consequence the use of real, finite-length, codes introduces a second source of inefficiency. The efficiency $\beta$ of slice reconciliation is given by:

\begin{equation}
\label{eq:slicebeta}
\beta = \frac{H(Q(X))-m+\sum_{i=1}^m R_i}{I(X;Y)}
\end{equation}

\begin{figure}
\centering
 \includegraphics[width=81mm]{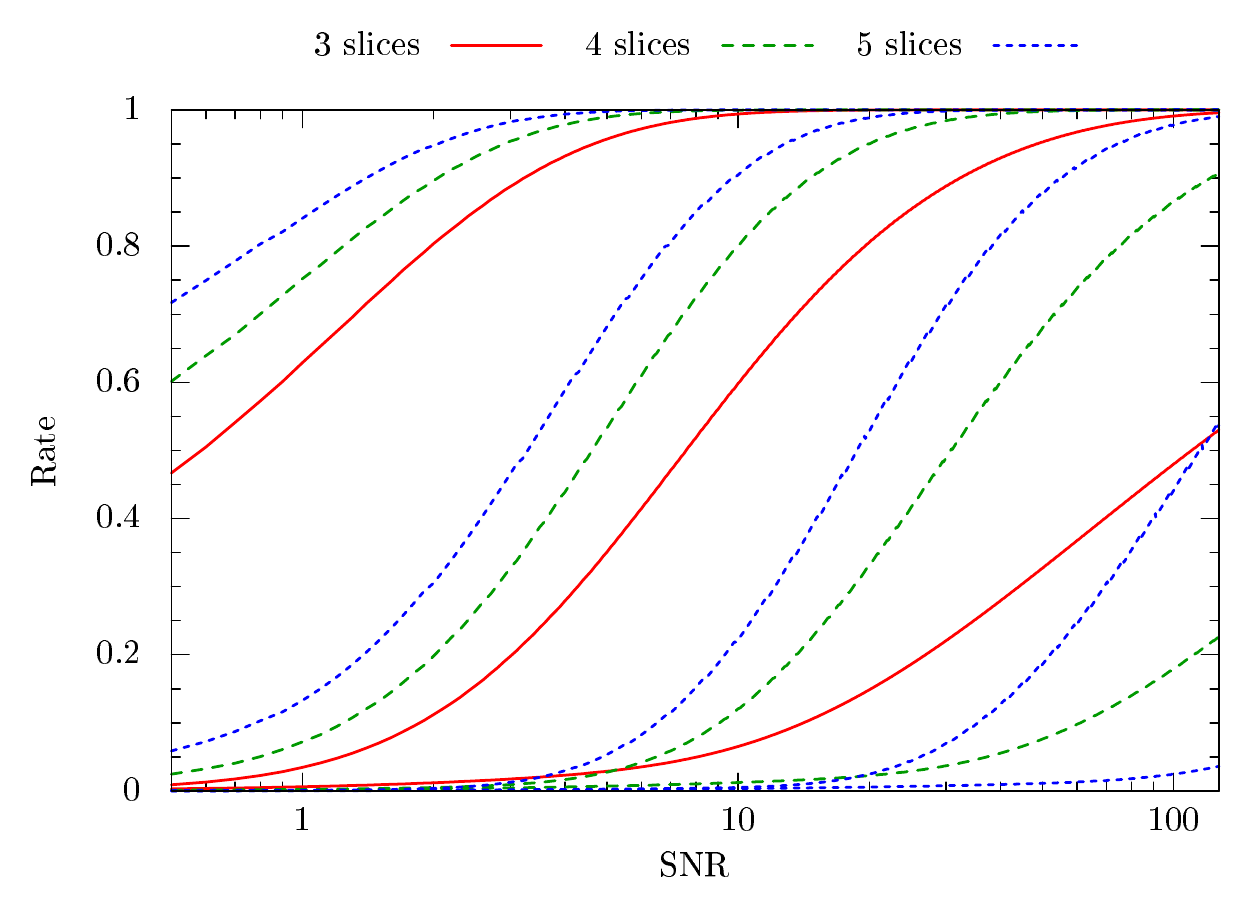}
  \caption{(Color online) Evolution of the value of the optimal rates for each slice with respect to the SNR for an optimal discretization of the real line into regular intervals. The lowest plots correspond to the least significant bits which are the noisiest bits. Solid red lines corresponds to 3 slices, dashed green lines corresponds to 4 slices, dotted blue lines corresponds to 5 slices.}
   \label{figure:optimumRatesConstantStep}
\end{figure}

Eq. \ref{eq:slicebeta} shows that $\beta$ is highly dependent in the rates of the available codes and how close they are to the channel capacities. For this reason we have chosen LDPC codes, well known for operating close to the capacity of symmetric binary input channels. The procedure is well known, for each rate the space of ensembles of codes is explored with an evolutionary algorithm \cite{SR:isit00} and for each ensemble the asymptotic behavior of the codes belonging to the ensemble can be evaluated with the Density Evolution algorithm \cite{CFRU:cl01}. The evolution of the value of the optimal rates for each slice with respect to the SNR for an optimal discretization of the real line into regular intervals is given in Fig. \ref{figure:optimumRatesConstantStep}. In practice, once the number of slices is fixed, for a given SNR we use Fig. \ref{figure:bestConstantStep} to choose the optimal quantization step and Fig. \ref{figure:optimumRatesConstantStep} to choose the optimal rates of the codes we need to design to decode the successive slices.

With optimal codes, the efficiency $\beta_{\text{disc}}$ of the discretization scheme is
\begin{equation}
\label{eq:slicebetadisc}
\beta_{\text{disc}} = \frac{H(Q(X))-m+\sum_{i=1}^m C_i}{I(X;Y)}
\end{equation}

\noindent where $C_i$ is the capacity of the channel corresponding to the $i$-th discretization layer. Assuming codes of efficiency $\beta_c < 1$ are used, the efficiency of the overall scheme is
\begin{align}
\beta &= \frac{H(Q(X))-m+\sum_{i=1}^m \beta_c C_i}{I(X;Y)} \notag \\
&= \beta_{\text{disc}} - (1-\beta_c) \gamma \label{eq:slicebetaloss}  \\
\text{ with } \notag\\
\gamma &=  \frac{\sum_{i=1}^m C_i}{I(X;Y)} \notag
\end{align}

The quantity $\gamma$ therefore controls the relationship between the lack of efficiency of individual error-correcting codes used and the efficiency loss that it causes on the slice reconciliation scheme. Because $H(Q(X)) \leq m$, when $\beta_{\text{disc}}$ is close to 1, $\gamma >1$. Typical values of $\gamma$ are between 1 and 2 as shown in Fig. \ref{figure:gamma}.

\subsection{Simulation Results}

\begin{figure}
\centering
 \includegraphics[width=81mm]{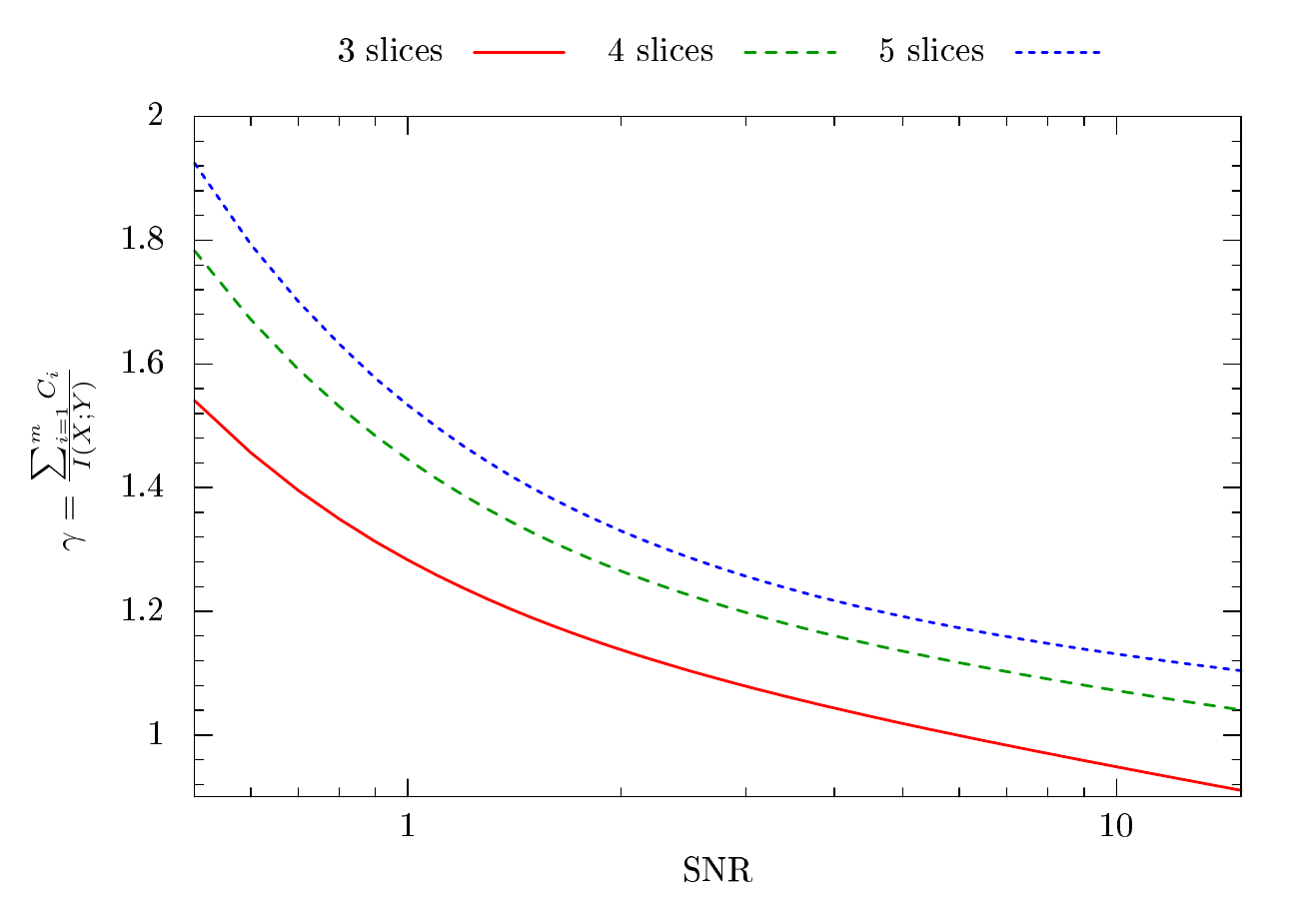}
  \caption{(Color online). Factor $\gamma$ indicating the sensitivity of slice reconciliation to the suboptimality of the error-correcting codes used for 3, 4 and 5 slices.}
   \label{figure:gamma}
\end{figure}

\begin{figure}
\centering
 \includegraphics[width=81mm]{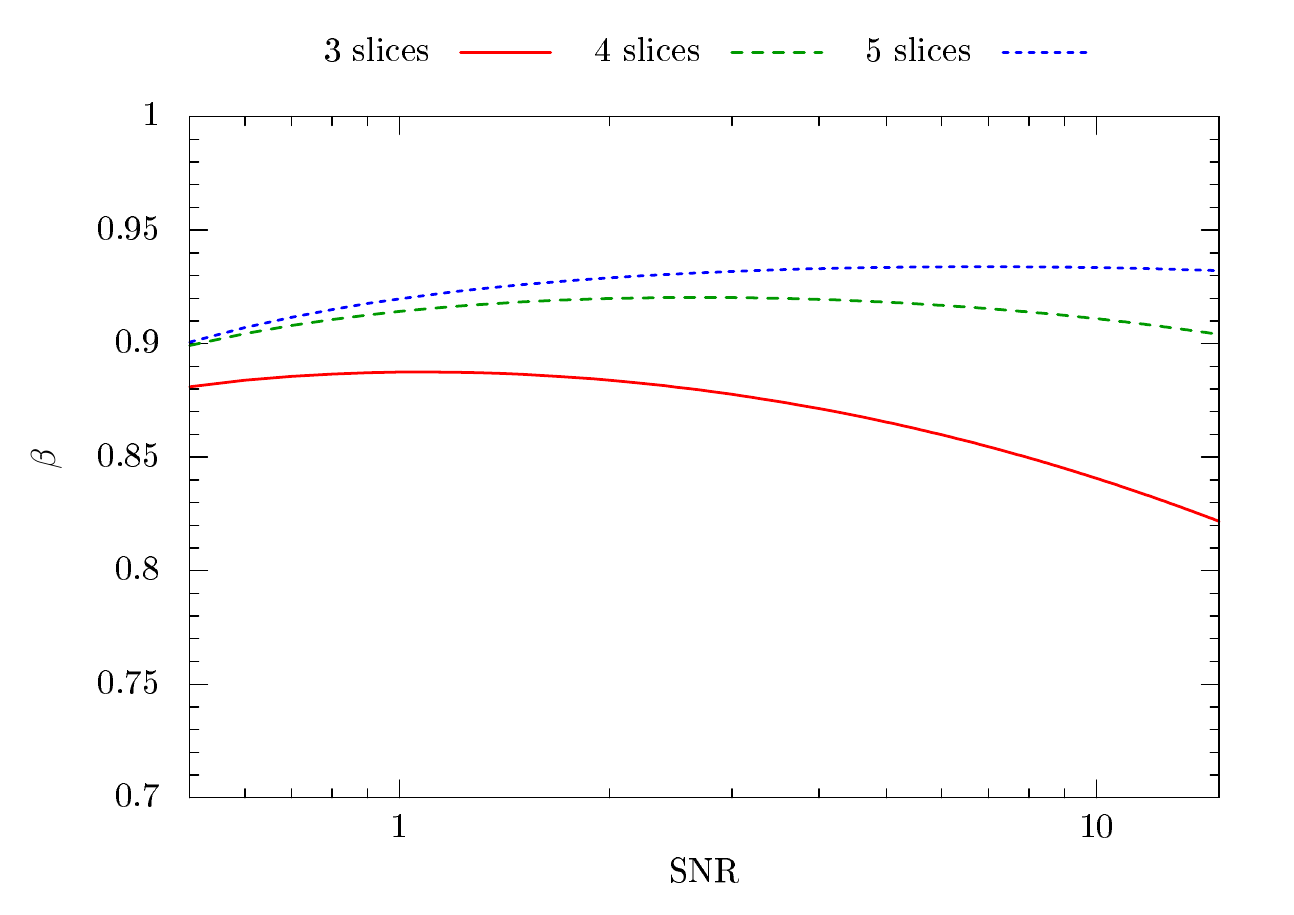}
  \caption{(Color online). Overall efficiency of slice reconciliation with error-correcting codes of efficiency $\beta_c=95\%$  for 3, 4 and 5 slices.}
   \label{figure:efficglobBeta095}
\end{figure}

\begin{figure}
\centering
 \includegraphics[width=81mm]{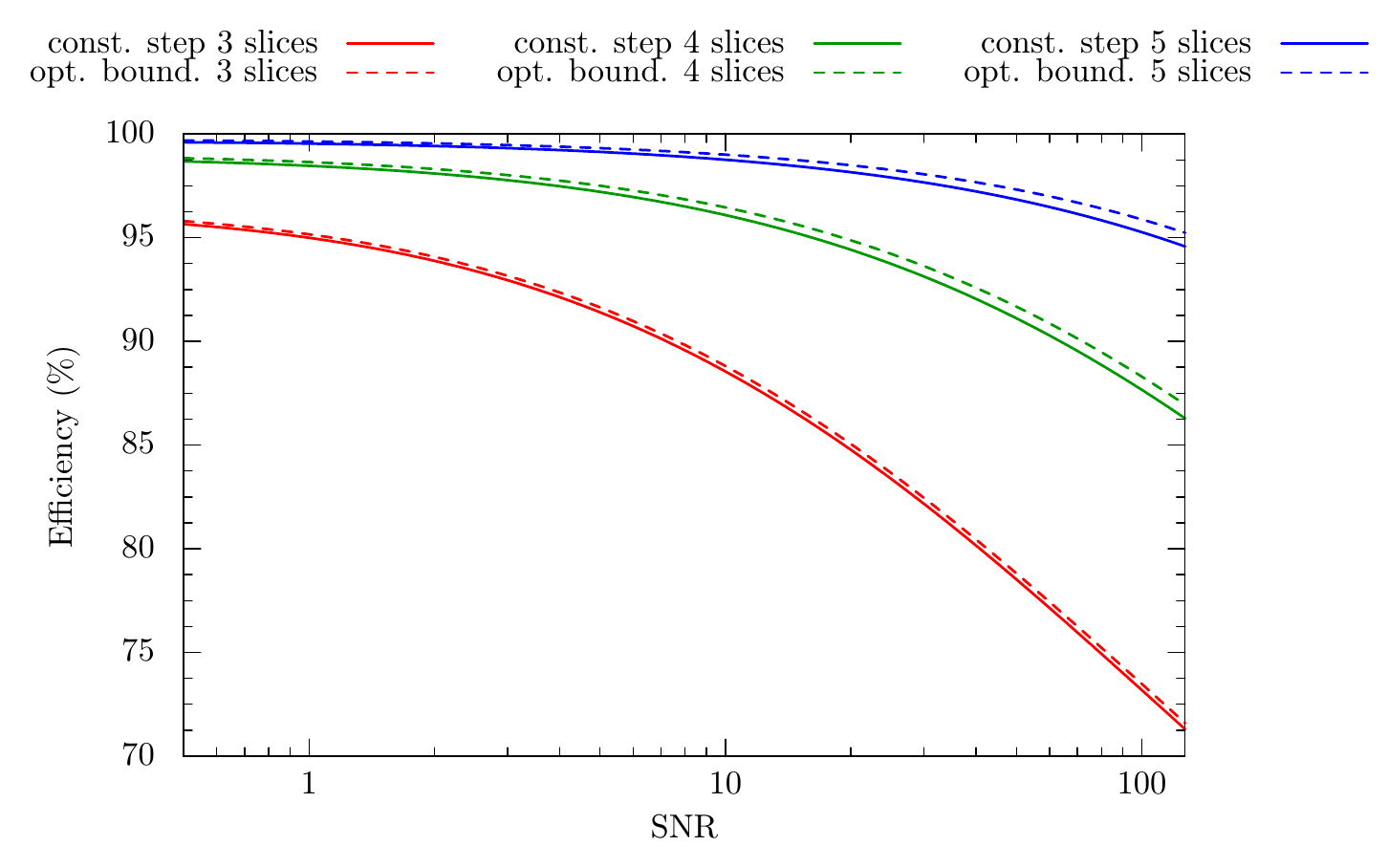}
  \caption{(Color online). Optimal quantization efficiency with respect to the SNR for a discretization of the real line into regular intervals (solid lines) and non regular intervals (dashed lines). Red lines give the discretization efficiency for 3 slices (8 intervals), green lines for 4 slices (16 intervals) and blue lines for 5 slices (32 intervals).}
   \label{figure:quantizationEfficiency}
\end{figure}

An optimization on the bounds of the discretization $Q(X)$ shows the following basic facts. For a fixed SNR, the higher the number of layers, the lower the discretization loss $I(X;Y) - I(Q(X);Y)$. It is always possible to make this loss negligible by increasing the number of layers. %, but then $H(Q(X))$ is farther and farther from $m$ in Eq. \ref{eq:slicebeta}, because $Q(X)$ is less and less uniform. 
This implies that $\gamma$ increases and can become much larger than 1 as shown in Fig. \ref{figure:gamma}. As seen in Eq. \ref{eq:slicebetaloss}, this means that adding layers requires error-correcting codes closer to the Shannon limit to minimize the loss on the scheme caused by the inefficiency of the individual codes. Overall, with codes having $\beta_c \geq 95\%$, the 5-slice scheme is the best on the SNR range $0.5-15$, and is always above 90\% efficiency, as shown on Fig. \ref{figure:efficglobBeta095}, thanks to its high quantization efficiency (see Fig. \ref{figure:quantizationEfficiency}), and despite its higher $\gamma$ value at low SNR. This is much better than results of \cite{LBG:pra07}, where an efficiency above 90\% could only be obtained for SNRs above 7. This is mainly due to the fact that we designed specific codes to decode each slice. Furthermore we perform error correction with codes of large length ($2^{20}$). 

As a summary, we show in the first two columns of Table \ref{tab:efficiencies} the best efficiencies obtained with slice reconciliation optimizing over the number of slices and the quantization step.  In the last two columns we show the efficiencies reported in \cite{JKL:pra11} with codes for the BIAWGNC. 

\begin{table}[h]
\caption{The first two columns show the efficiencies achieved with slice reconciliation with respect to the SNR. The last two columns show the efficiencies achieved with multi-edge LDPC codes with respect to the SNR, these values were reported in \cite{JKL:pra11}. }
\centering
\begin{tabular}{c|c|c|c}
%\label{tab:efficiencies}
\multicolumn{2}{c|}{AWGN} & \multicolumn{2}{c}{BIAWGN}\\
\hline
SNR     & Efficiency & SNR & Efficiency \\
\hline
%0.5      & 90.5\% &  \\
0.55     & 93.4\% & 0.0075 & 95.9\% \\
0.86     & 93.7\% & 0.0145 & 96.6\% \\
1         & 94.2\% & 0.029   & 96.9\% \\
3         & 94.1\% & 0.075   & 95.8\% \\
5.12    & 94.4\% & 0.161   & 93.1\% \\
14.57   & 95.8\% & 1.097   & 93.6\% \\
66.10   & 94.8\% & &
%156.25  & 90.1\% & & 
\label{tab:efficiencies}
\end{tabular}
\end{table}

For SNRs below 0.5, the multidimensional methods of \cite{JKL:pra11} are more competitive than slice reconciliation. Indeed in that case $I(X,Y) \ll 1$, and the main limitation of multidimensional methods that they can only extract 1 bit per pulse is not a problem. Therefore the combination of multidimensional methods and slice reconciliation with up to 5 slices yields an efficiency above 90\% for SNRs ranging from 0.01 to 100.
For SNRs above 10, the capacity of the highest layer is sufficiently close to 1 to be able to use a simple, fast, hard decoding code such as a BCH code to decode it. As an alternative, it is always possible to use a code in a regime of higher SNR than its initial threshold SNR. In this case the following efficiency can be obtained:
\begin{equation}
\beta_s=\beta_{s_0} \frac{\log (1+s_0)}{\log (1+s)}
\end{equation}
where $s,\beta_s$ denotes the target SNR and $s_0,\beta_{s_0}$ the original SNR and efficiency \footnote{Let $R$ be the rate of a binary code that is used for reconciliation with $s_0$ as SNR, then $\beta_{s_0}=R/\log(1+s_0)$. The same code can be used for reconciliation with a lower amount of noise, in this case the efficiency is simply given by:  $\beta_{s}=R/\log(1+s)=\beta_{s_0} \log(1+s_0)/\log(1+s)$.}.

At the other end of the spectrum, low-rate slices are decoded with multi-edge LDPC codes which can have an efficiency above 95\% for rates $0.1 - 0.02$  \cite{JKL:pra11}. For even lower rates, multi-edge LDPC codes can be combined with a length $k$ repetition code without a significant efficiency loss \cite{LG:arxiv10}:
\begin{equation}
\beta_s = \beta_{s_0} \frac{s\log (1+s_0)}{s_0\log (1+s)} \underset{s_0 \approx 0}{\approx} 1 
\end{equation}
where $s=s_0/k$. Alternatively, the slices can be fully revealed. Revealing a lower slice is not equivalent to reducing the number of slices, since the knowledge of the lowest slices helps the soft decoding of the upper slices.

\section{Application to High Bit Rate CVQKD}
\label{sec:application}
\begin{figure}[h!]

\centering
\begin{overpic}[scale=0.6,unit=1mm]{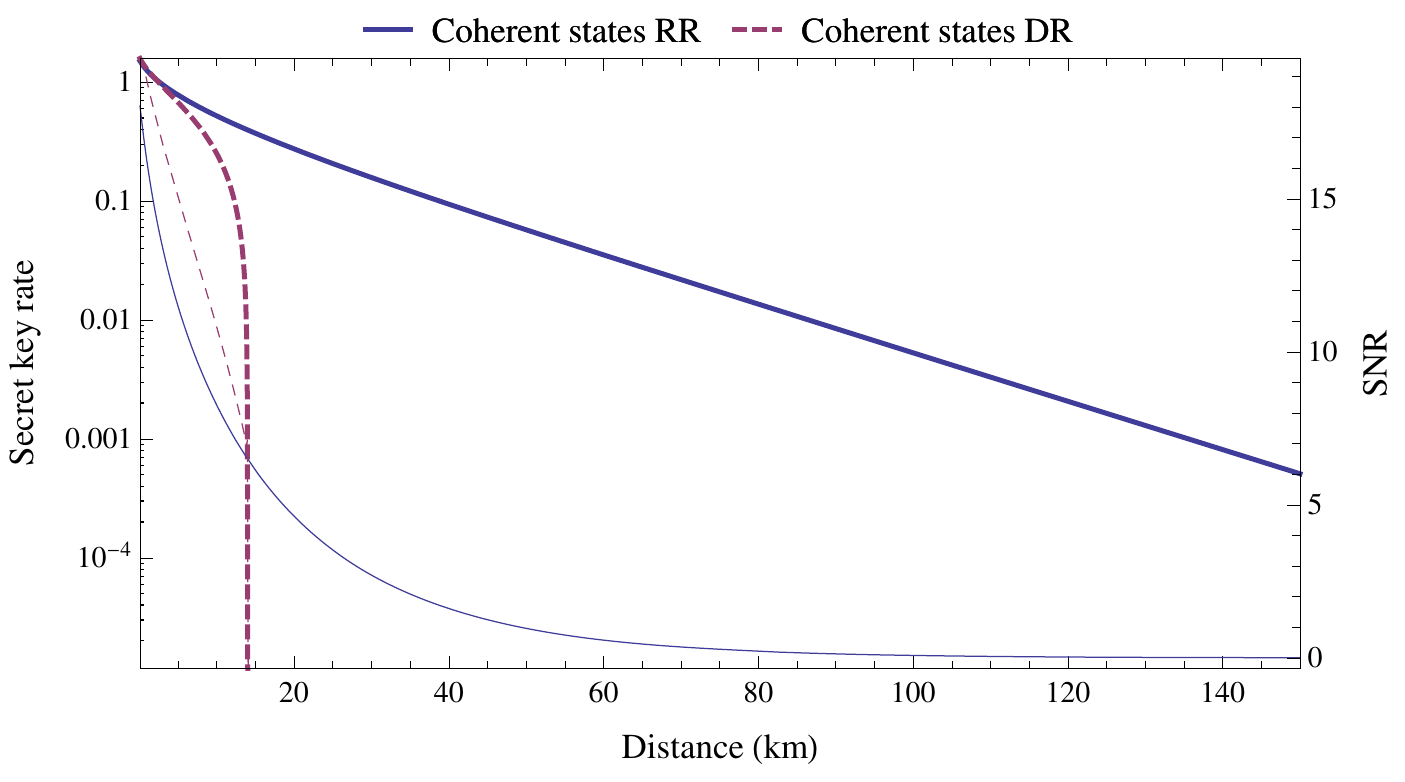}
\put(61,30){\includegraphics[scale=.21]%
{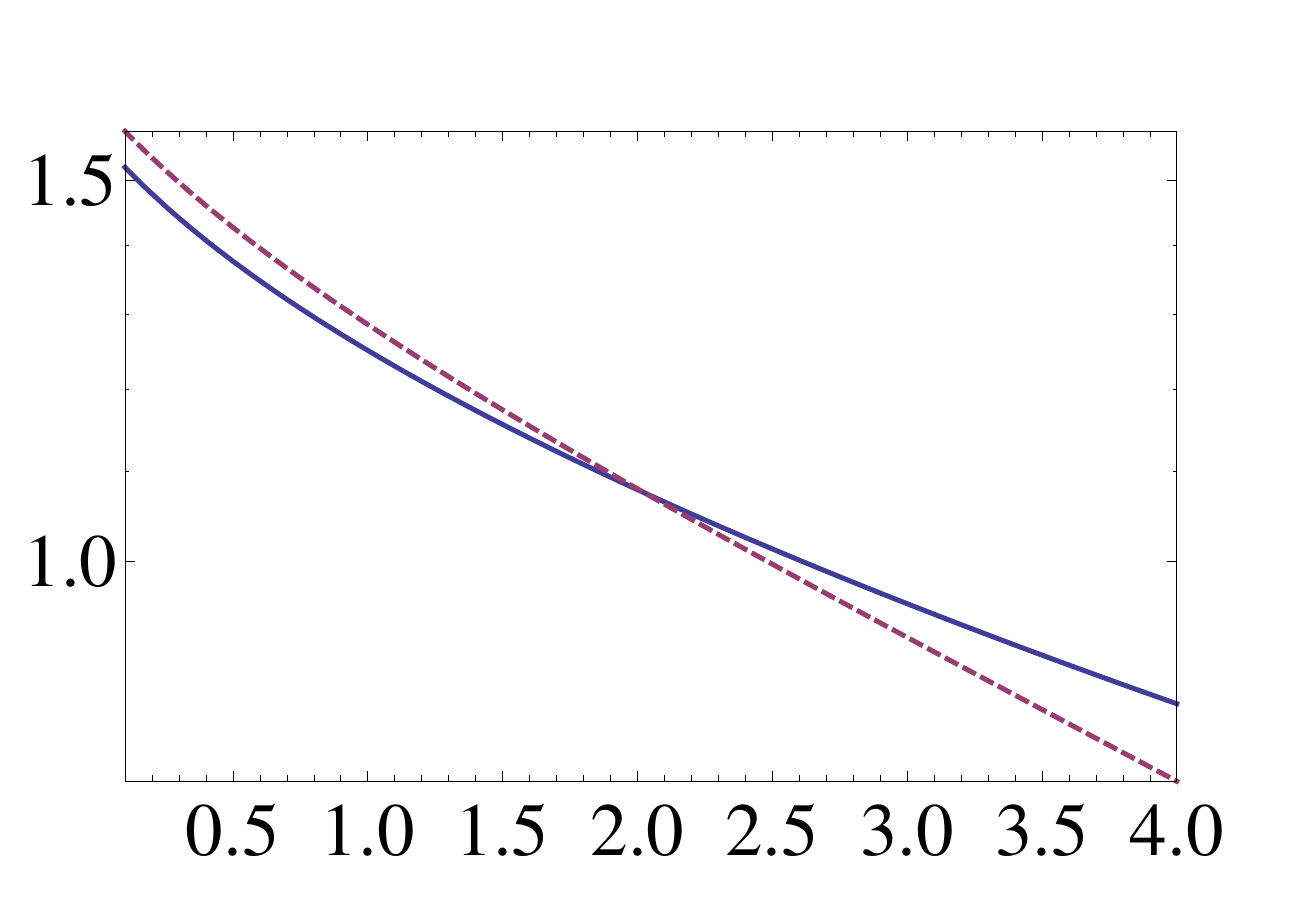}}
\end{overpic}
\caption{(Color online). Comparison of the DR and RR secret key rate with ideal measurement devices. The thick lines show secret key rate, the thin lines the optimal SNR which is equal to $V_A/T$ up to some small excess noise related term. The secret key rate of the first four km is zoomed in the upper right corner. Parameters: $\alpha=0.2,\xi=0.0015V_A$.}
\label{fig:pfctSkr}
\end{figure}

For all our simulations, we have computed the secret key rate against collective attacks \cite{GC06, NGA06}, which is equivalent to the secret key rate against general attacks in the limit of large block lengths. When considering finite size effects \cite{LGG:pra10, JKDL:pra12}, the performance of reconciliation is not affected but the modulation variance that yields the optimal key rate is different than in the asymptotic case; the secret key rate is also lower in this scenario than in the asymptotic one at any distance, partly because the estimated value of the excess noise is increased to take into account the statistical uncertainty of the estimator. The secret key rate greatly varies between the direct and reverse reconciliation scenarios. In Fig. \ref{fig:pfctSkr} we plot both scenarios with parameters $\xi=0.0015 V_A,\alpha=0.2$, where $V_A$ is the variance of Alice's input signal and ideal measurement devices and $\alpha$ is the loss coefficient of the optical fiber. For distances shorter than 2 km, DR is a better option but the curve drops sharply and reaches zero before 15 km which corresponds to the DR limit of 3dB. RR on the other hand has no theoretical limitation and with the chosen parameters at 100 km still yields a secret key rate close to $5\cdot 10^{-3}$ bits per symbol. These secret key rates are the maximized rates over the variance of Alice's input signal. The corresponding SNR values are plotted with the same pattern and colour as the corresponding secret key rate with smaller width. The remaining figures in this section follow the same convention. 

\begin{figure}[h]
\centering
\begin{overpic}[scale=.6,unit=1mm]{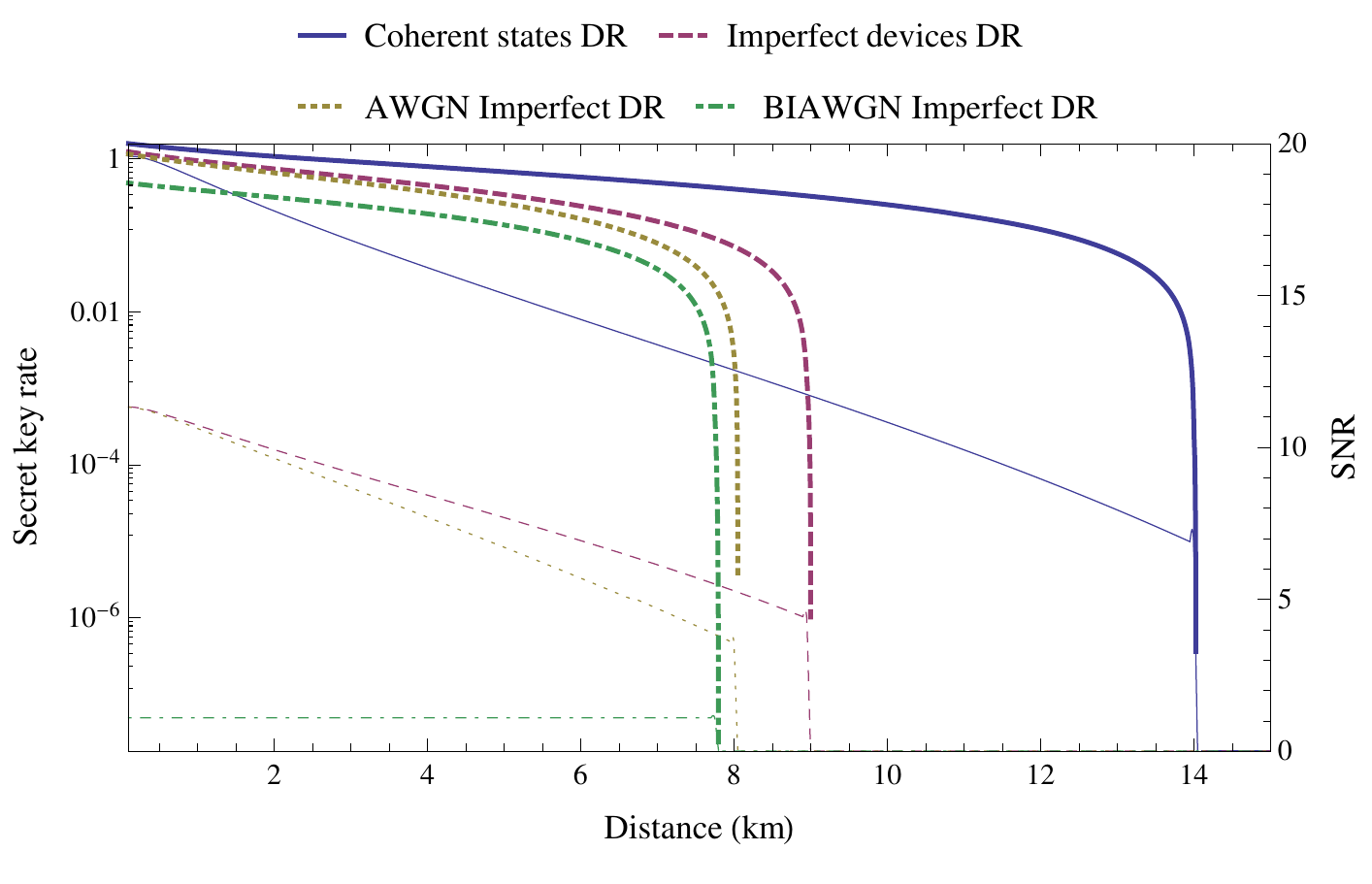}
\end{overpic}
\caption{(Color online). Secret key rate in the DR scenario. From top to bottom the curves show the secret key rate with: ideal measurement devices, realistic devices characterized by a finite detection efficiency $\eta$ and an electronic noise $v_{elec}$, slice reconciliation and reconciliation over a BIAWGN. The thick lines show secret key rate, the thin lines the optimal SNR which is a function of the input signal variance. Parameters: $\xi=0.0015 V_A,\alpha=0.2,\eta=0.6,v_{elec}=0.01$.}
\label{fig:skrDR}
\end{figure}

The optimization of the quantization step allows to increase the secret key rate in the short distance regime. This is particularly noticeable in the DR scenario. In Fig. \ref{fig:skrDR} we show the achievable secret key rate with ideal measurement devices. We have chosen three scenarios for comparison: 1) imperfect detection devices and perfect reconciliation 2) slice reconciliation and 3) reconciliation over a BIAWGN of the same SNR (limit case of the multidimensional channels \cite{JKL:pra11}). The four curves run separated over the whole region considered, the main reason is that the optimal $V_A$ values correspond to high SNR values (plotted in the same curve) which translates into an advantage for slice reconciliation. We would like to highlight that for very short distances, slice reconciliation allows to distill for the first time more than one secret bit per channel use.

\begin{figure}[h]
\centering
\begin{overpic}[scale=.6,unit=1mm]{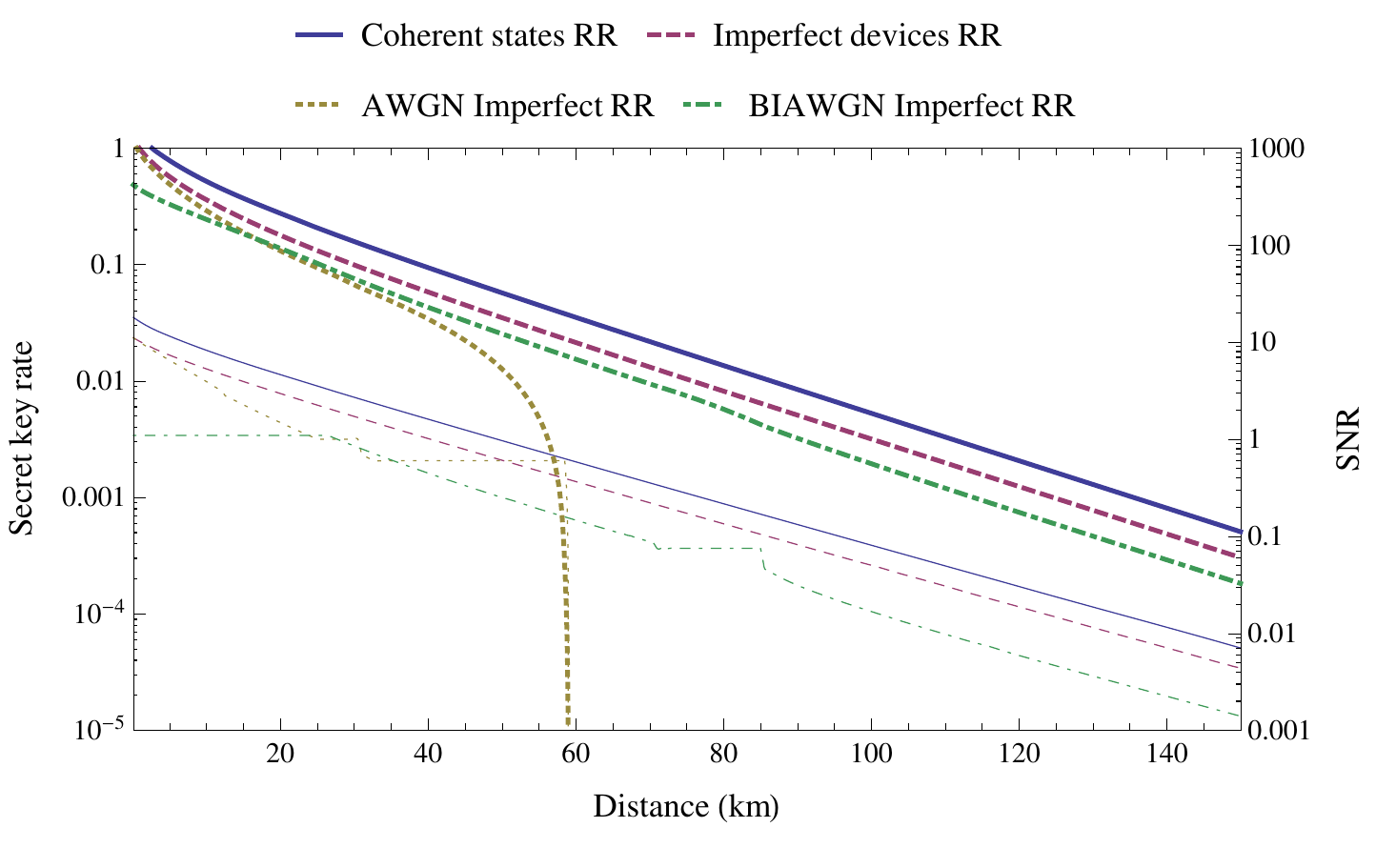}
\put(58.5,32){\includegraphics[scale=.23]%
{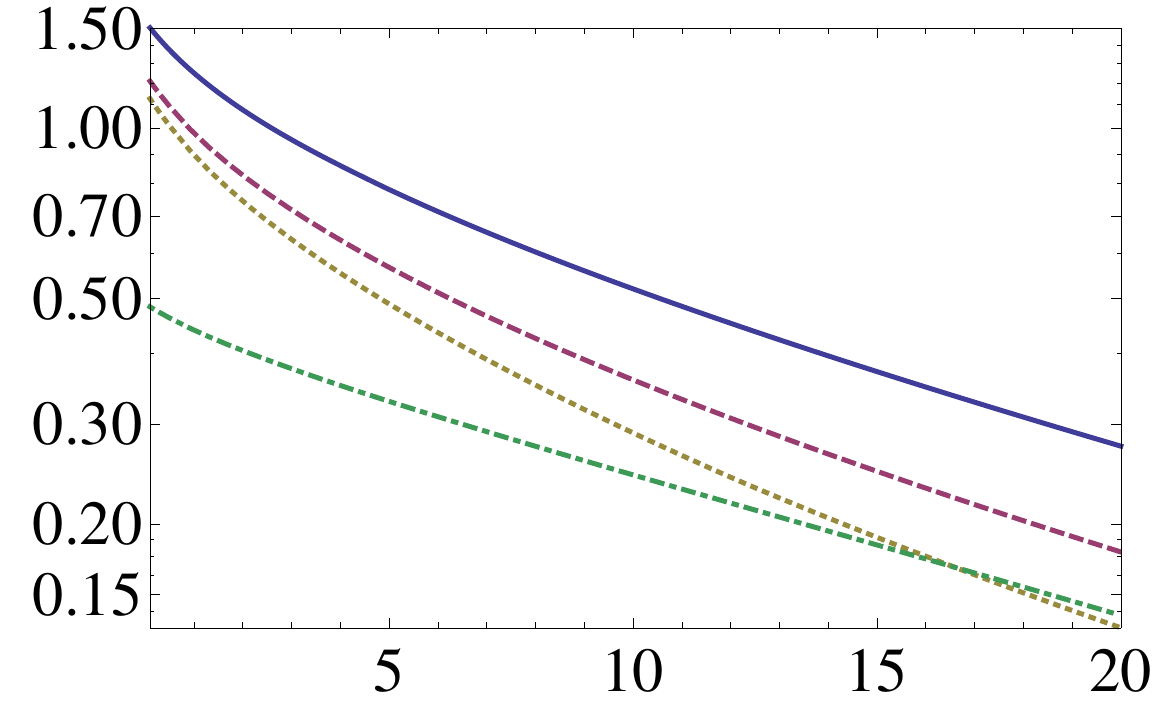}}
\end{overpic}
\caption{(Color online). Secret key rate in the RR scenario. From top to bottom the curves show the secret key rate with: ideal measurement devices, realistic devices characterized by a finite detection efficiency $\eta$ and an electronic noise $v_{elec}$, slice reconciliation and reconciliation over a BIAWGN. The thick lines show secret key rate, the thin lines the optimal SNR which is a function of the input signal variance. The secret key rate of the first twenty km is zoomed in at the upper right corner. Parameters: $\xi=0.0015 V_A,\alpha=0.2,\eta=0.6,v_{elec}=0.01$.}
\label{fig:skrRR}
\end{figure}

In the reverse reconciliation scenario the advantage of our implementation of slice reconciliation is limited to distances below $13$ km. The reason lies in the increasing difficulty of optimizing multilevel coding schemes for low SNRs. Furthermore, binary encodings are optimal in the low SNR regime. The reason is that the capacity of the associated channel, the BIAWGN, converges to the capacity of the AWGN channel as the SNR goes to zero. In fact, binary encodings have successfully been used for long distance CVQKD \cite{LAB:pra08}. We observe this behaviour in Fig. \ref{fig:skrRR}: below $13$ km there is an advantage in using slice reconciliation, but over this distance binary encodings lead and allow to distill secret key over large distances \cite{JKL:pra11}. 

We used the experimental system reported in \cite{JKL+:natphoton13} and operated it in the high SNR regime for very low losses between Alice and Bob. For a SNR of 19 and a line transmission of 0.995, we obtained an excess noise of 0.03 shot noise units (SNU) on Bob's side, i.e. an excess noise of 0.05 SNU on Alice's side for a measured homodyne detection efficiency of 0.6 and an electronic noise of 0.01 SNU. We obtained a practical reconciliation efficiency of 95\% and the secret key rate per pulse is about 1.02 in the reverse reconciliation scenario while it reaches 1.04 in the direct reconciliation scenario. These measurements confirm the possibility to extract more than one secret bit per pulse with a CVQKD system.

\begin{figure}
\centering
 \includegraphics[width=81mm]{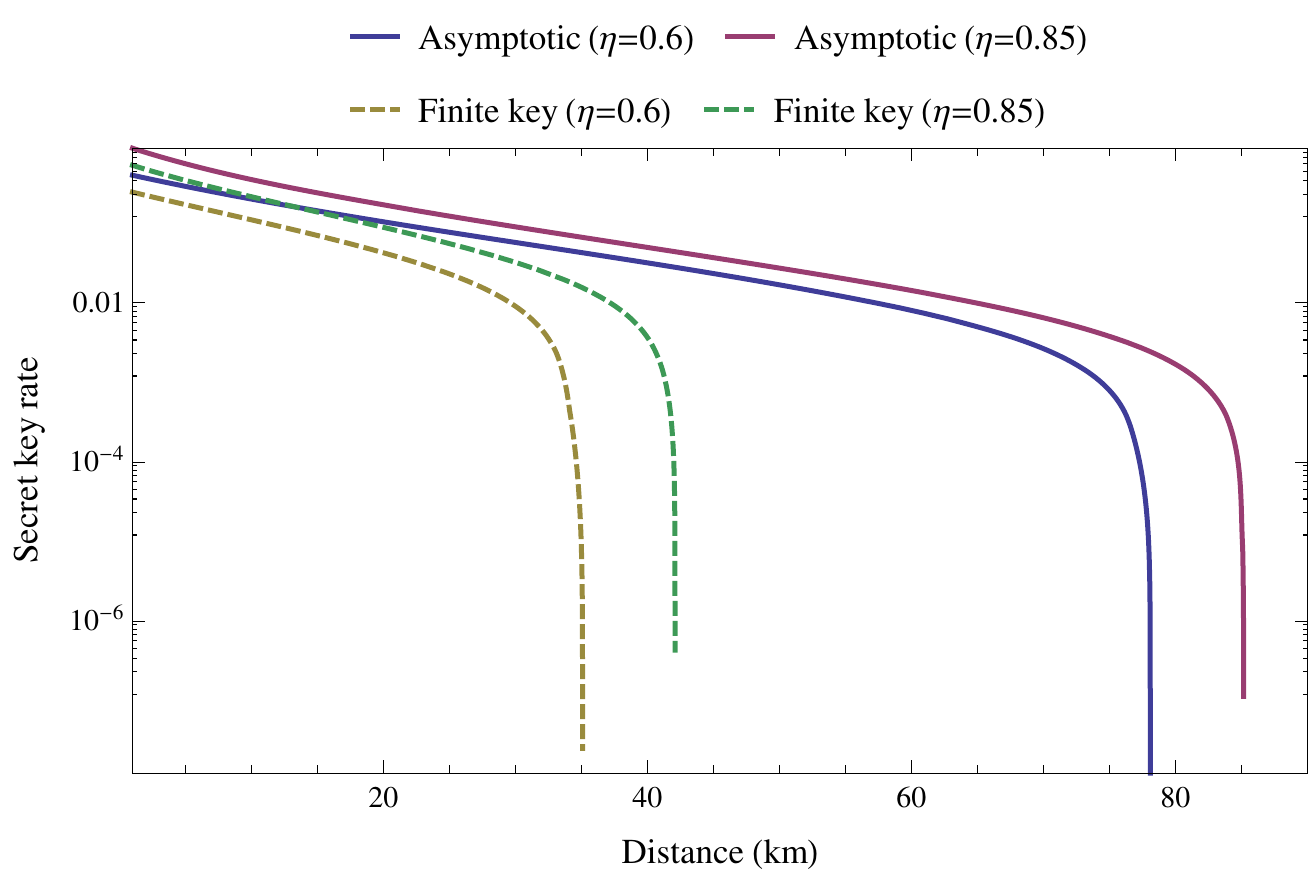}
  \caption{(Color online) Secret key rate in the RR scenario with a heterodyne detection and the security proof of \cite{Lev:arxiv14}. From top to bottom the curves show the secret key rate in the asymptotic and finite key scenario for realistic devices characterized by an electronic noise $v_{elec}=0.001$ and a finite detection efficiency $\eta=0.85$ and $\eta=0.6$. We consider here the paranoid mode where the noise added by the detection can be manipulated by the attacker. The solid lines show the asymptotic secret key rate, the dashed lines show the secret key rate with finite blocks of length $n=10^9$. Other parameters: $\xi=0.0015 V_A,\alpha=0.2$, blocks size $n=10^9$ and the security parameter $\epsilon=10^{-10}$.}
   \label{figure:skr_finite_key}
\end{figure}

\begin{figure}
\centering
 \includegraphics[width=81mm]{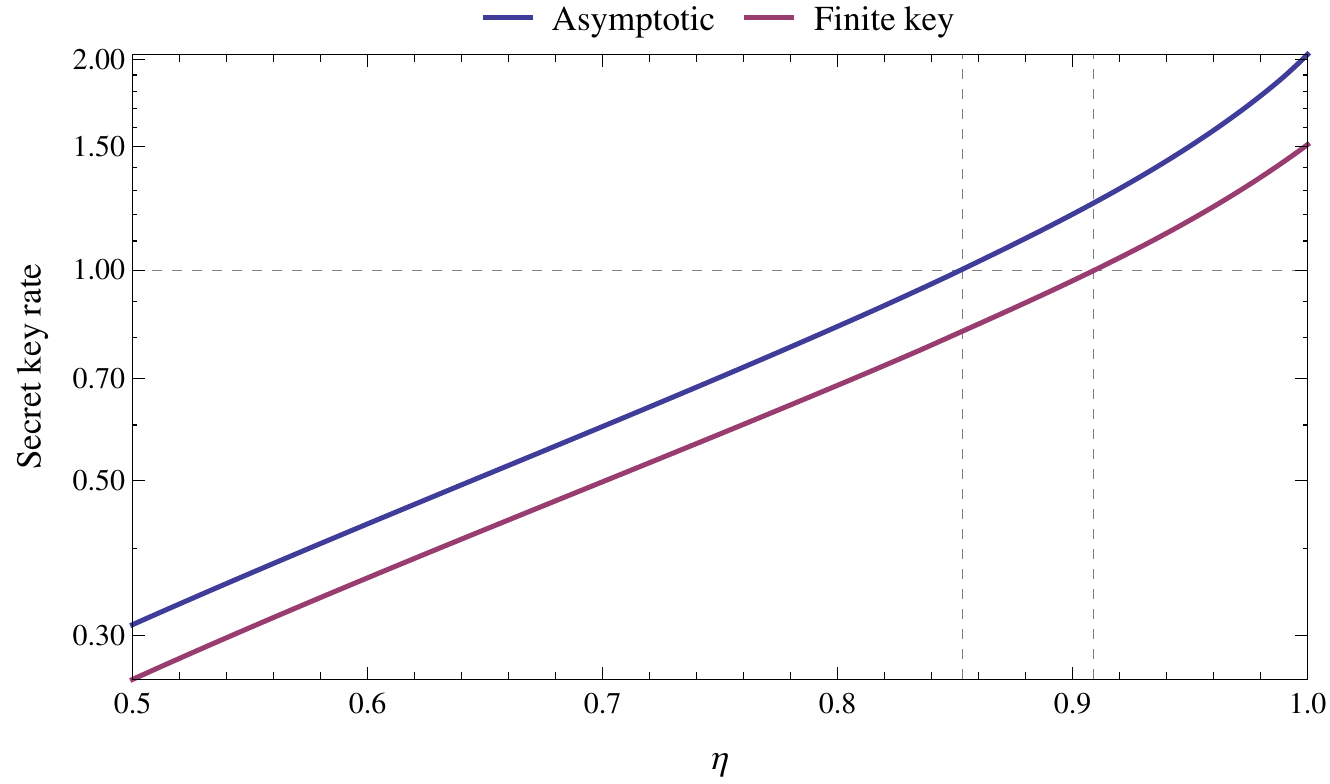}
  \caption{(Color online) Secret key rate in the RR scenario with a heterodyne detection and the security proof of \cite{Lev:arxiv14}. From top to bottom the curves show the secret key rate in the asymptotic and finite key scenario for realistic devices characterized by an electronic noise $v_{elec}=0.001$ with respect to the detection efficiency. We consider here the paranoid mode where the noise added by the detection can be manipulated by the attacker. Other parameters: distance $d=0.1km$, $\xi=0.0015 V_A,\alpha=0.2$, blocks size $n=10^{10}$ and the security parameter $\epsilon=10^{-10}$.}
   \label{figure:skr_finite_key_eta}
\end{figure}

We investigated the robustness of these results in the composable security framework presented in \cite{Lev:arxiv14}. In the same way than our previous simulations, we optimized the secret key rate with respect to the reconciliation efficiency and considered both direct and reverse reconciliation scenarios with imperfect devices. However, we considered the heterodyne protocol, as described in \cite{Lev:arxiv14}, in the paranoid mode where the imperfections of the detector are assumed to be controlled by Eve and in the limit of finite-length data blocks. This corresponds to the most secure known scenario and as expected the secret key rate is lower than in our previous simulations as shown in Fig. \ref{figure:skr_finite_key}. 
With a heterodyne detection characterized by an efficiency $\eta=0.6$ and an electronic noise $v_{elec}=0.01$, the secret key rate vanishes at about 30 km. This is why we plot in Fig. \ref{figure:skr_finite_key} the secret key rate in both the finite key and the asymptotic scenario for realistic improvements of the heterodyne detection. All the curves are plotted with an electronic noise $v_{elec}=0.001$ which is achievable with cooled heterodyne detections. With a heterodyne detection efficiency of $60\%$ a secure distance of about 35 km can be achieved in the finite key scenario while an improved heterodyne detection efficiency of $85\%$ would allow us to exchange keys at about 80 km but in the asymptotic limit. One can see that the secret key rate drops below 1 bit per symbol with a heterodyne detection efficiency of $60\%$. We show in Fig. \ref{figure:skr_finite_key_eta} that it is still possible to exchange secret keys with a rate higher than 1 bit per symbol at short distance ($0.1$ km) even in the paranoid mode and using finite data blocks but at the expense of improving the detection efficiency to about $91\%$ and using data blocks of size $10^{10}$.

In Table \ref{tab:skr-comparison} we compare a recent DVQKD experiment yielding high secret key throughput \cite{CFL+:apl2014} with the two CVQKD scenarios depicted in Fig. \ref{figure:skr_finite_key}. Columns two and four correspond to secret key rate per signal, while columns seven to nine correspond to secret key throughputs. Columns three and five respectively give the ratios between columns two and six and between columns four and six. In order to get a throughput figure, we multiply the secret key rates by the corresponding clock rate. Column seven corresponds to a clock rate of 1 MHz as reported in \cite{JKL+:natphoton13}, while column eight reports the expected throughput for a reasonable improvement of the clock rate to 50 MHz.

On the hardware side, increasing the clock rate to about 50 MHz is not a big deal: high bandwidth optical modulators and acquisition cards are commercially available while homodyne detections running at a few hundreds MHz have already been reported \cite{HJW+:qcrypt13}. As regards the post-processing, privacy amplification can be done at a few hundreds of MHz on one core of a modern CPU but high efficiency error correction as described in this paper would require at least one modern GPU and probably two. More generally, when dealing with continuous values at such speeds, every step, such as random numbers generation and network communication, must be implemented carefully.

\begin{table*}
\centering
\begin{tabular}{c|c|c|c|c|c|c|c|c}
      & \multicolumn{5}{c|}{rate} & \multicolumn{3}{c}{throughput} \\
      & CVQKD$^1$ & ratio$^1$$^/$$^3$ & CVQKD$^2$ & ratio$^2$$^/$$^3$ & DVQKD$^3$ & 1 MHz$^1$ & 50 MHz$^2$ & 1 GHz$^3$ \\
\hline
100 m	& 2.7E-01	& 17 & 6.0E-01 & 39 & 1.5E-02 &	2.7E+05 & 	3.0E+07 & 	1.5E+07 \\
10 km	& 1.1E-01	& 12 & 2.2E-01 & 23 & 9.5E-03 &	1.1E+05 & 	1.1E+07 & 	9.5E+06 \\
30 km	& 9.0E-03	& 2 & 3.2E-02 & 9 & 3.6E-03 & 	9.0E+03 & 	1.6E+07 & 	3.6E+06 \\
40 km	& -	           & -            & 3.7E-03 & 2 & 2.2E-03 &	-            & 	1.8E+05 & 	2.2E+06 
\end{tabular}
\caption{Comparison of CVQKD and DVQKD. $^1$ corresponds to a realistic setting characterized by an electronic noise $v_{elec}=0.001$, a detection efficiency $\eta=0.6$ and a 1MHz repetition rate \cite{JKL+:natphoton13}. $^2$ corresponds to an optimistic setting characterized by an electronic noise $v_{elec}=0.001$, a detection efficiency $\eta=0.85$ and a 50MHz repetition rate. $^3$ corresponds to the DVQKD data reported in \cite{CFL+:apl2014}. For CVQKD, the figures are obtained using the security proof of Ref. \cite{Lev:arxiv14} as in Fig. \ref{figure:skr_finite_key}.}
\label{tab:skr-comparison}
\end{table*}

\section{Conclusion}
We have optimized the performance of practical reconciliation schemes for CVQKD, and the resulting schemes have above 90\% efficiency for any SNR, which leads to higher key rates than those reported in past CVQKD experiments \cite{JKL+:natphoton13}. Notably, for distances below $100$m, more than 1 bit per symbol can be distilled. The expected throughput with a CVQKD clock rate of 1 MHz, as reported in \cite{JKL+:natphoton13}, is lower than the best DVQKD reported throughput, which uses a 1 GHz clock rate \cite{CFL+:apl2014}. However, we predict (see Table \ref{tab:skr-comparison}) that reasonable improvements of the CVQKD hardware would result in throughputs higher than those of DVQKD in distances up to 30 Km. 

\section{Acknowledgements}
This research was supported by the French National Research Agency, through the HIPERCOM (2011-CHRI-006) project, by the DIRECCTE Ile-de-France through the QVPN (FEDER-41402) project, and by the European Union through the Q-CERT (FP7-PEOPLE-2009-IAPP) project.

DE acknowledges financial support from the European CHIST-ERA project CQC (funded partially by MINECO grant PRI-PIMCHI-2011-1071).

\end{document}